# Time-delay signature concealed broadband gain-coupled chaotic laser with fiber random grating induced distributed feedback

Yanping Xu[1]**, Liang Zhang[1]**, Ping Lu[2], Steven Mihailov[2], Liang Chen[1] and Xiaoyi Bao[1]*

[1]Department of Physics, University of Ottawa, 25 Templeton Street, Ottawa, ON, K1N 6N5, Canada

[2]National Research Council Canada, 100 Sussex Drive, Ottawa, Ontario, K1A 0R6, Canada

*Corresponding author: xbao@uottawa.ca;

**These authors contributed equally to this work

**Abstract**: A broadband chaotic laser with a flat power spectrum extending up to 8.5GHz is achieved by injecting continuous wave laser light into a chaotic diode laser perturbed by fiber random grating induced distributed feedback, which forms a gain-coupled chaotic laser system. More than triple bandwidth enhancement is realized with appropriate frequency detuning between the master and slave lasers. Unlike normal chaotic lasers, the output from such a broadband chaotic laser is free from time-delay signature, which can be applied for practical real-time random number generations and will find useful applications in the fields of information security and computation systems.

## 1. Introduction

Broadband chaotic semiconductor lasers have been extensively studied during the past several decades for their valuable applications in secure communications [1-5], random number generations [6-12], chaotic lidar [13,14], compressive sensing [15], and time domain reflectometry [16,17]. It is well-known that semiconductor lasers exhibit a rich variety of lasing dynamics with external perturbations such as optical feedback, laser injection, current modulation, and optoelectronic feedback [18-20]. Among these lasing dynamics, chaotic oscillation in semiconductor lasers has drawn intensive attentions, which has been realized by various perturbation techniques. Perturbations can overcome the limited bandwidth of the intrinsic chaotic intensity oscillations in semiconductor lasers dominated by laser relaxation oscillation and significantly broaden the chaotic bandwidth of the laser output. Among the perturbation techniques, the passive optical feedback and active laser injection configurations are the most effective and simplest schemes for chaotic oscillation generation and bandwidth enhancement [21]. However, the optical feedback based chaotic laser systems often suffer from the time-delay signature (TDS) in chaotic outputs, which exhibits as periodicities in output time series. This is induced by the photon round-trip in the external cavity formed by the optical feedback. The TDS in chaos communications will allow eavesdroppers to crack the encryption systems and deteriorate the security of the communication systems. Various feedback configurations have been proposed and investigated to overcome the TDS in chaotic laser output, including single mirror feedback [22], dual-path feedback from double mirrors [23], fiber Bragg grating (FBG) feedback [24,25], feedback with cascaded coupled laser injection [26], on-chip integrated optical feedback [27] and so on. In FBG feedback configuration [24,25], the laser frequency has to be tuned to edges of the main lobe of FBG reflection spectrum, which leads to a significant power loss in reflection, weakening the feedback strength to the semiconductor laser. The FBG spectrum that is sensitive to environmental disturbances will also bring instability to the detuning frequency and compromise the TDS suppression. Furthermore, the broadening of the chaotic bandwidth based on only the optical feedback perturbation is still limited by the intrinsic relaxation oscillation of the single semiconductor laser. Thus it cannot meet the high-speed requirement for the current fast chaos communication systems. Laser injection technique has also been combined with the optical feedback technique to further enhance the chaotic bandwidth [28, 29], however, TDS still remains in the previously reported works and therefore the proposed chaotic laser systems cannot function in real-time mode for random bit generations.

In this paper, we experimentally demonstrate a broadband gain-coupled chaotic laser system with random distributed feedback from a fiber random grating and additional laser injection. The TDS in the output signal of the chaotic laser is completely suppressed thanks to the fiber random grating feedback, which significantly complicates the external feedback cavity features and thus the chaotic laser dynamics. The random distributed feedback erases the original cavity modes in both slave and master lasers, and creates larger numbers of random cavity modes in the new chaotic laser system due to the randomness feature of the distributed feedback. As a result, different chaotic lasing dynamics are achieved comparing with normal chaotic laser. The proposed broadband chaotic laser is believed to be an excellent candidate for realizing physical random number generators that can function in real time and offer random bit streams with high-quality randomness.

## 2. Experiments and results

The experimental configuration of the proposed broadband chaotic laser system is shown in Fig. 1(a). Two distributed feedback coaxial laser diodes (LDs) are used as light sources, one as slave LD and the other as master LD. The output power of the LD as a function of the driving current is shown in Fig. 1(b). The measured threshold driving current is about 6mA, above which the LD output power increases linearly with the driving current. The optical spectrum of the LD is measured as shown in Fig. 1(c), where the center wavelength is located around 1548nm. The spectral peak of the LD could be adjusted through temperature control. Light from the slave LD passing through CIR 1 was first amplified by an EDFA and PC 1 was used to adjust the state of polarization (SOP) of the intra-cavity light. The amplified light was then sent through the CIR 2 to be scattered back by a fiber random grating. The randomly backscattered light was passed through Atn 1 and guided back to the slave LD after being combined with the light from the master LD. The fiber random grating sample was fabricated through the plane-by-plane inscription without phase control [30]. A total of ~25000 index modified planes with random spatial interval from 0 to 3.5 μm and random index modulations were introduced, which act as enhanced Rayleigh scattering centers randomly distributed along the 2.5cm long standard single mode fiber. The enlarged schematic image in Fig. 1(a) exhibits that multiple interferences occur along the fiber random grating with reflections from different spots. The reflectivity of the fiber random grating was measured to be ~0.001 over a broad spectral range from 1540nm to 1560nm. It is worth mentioning that the transmission loss of the random grating sample was controlled to be less than 3dB, which enables the direct detection of the chaotic light at the output end of the grating sample. A 12GHz bandwidth photodetector (1544-A, New Focus) was used to monitor the chaotic laser output, which was afterwards digitized by an oscilloscope (DS081204B, Agilent). An optical spectral analyzer (AQ6375, Yokogawa) and an electrical spectral analyzer (E4446A, Agilent) were also used to characterize the laser output spectrum.

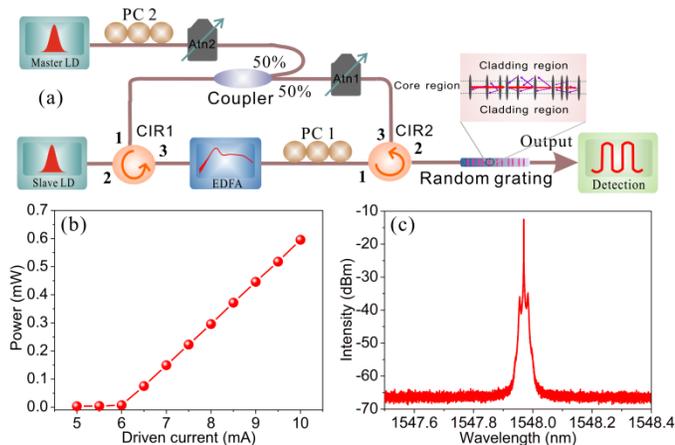

Fig. 1. (a) Experimental setup of the proposed broadband chaotic laser system. Atn: Attenuator, CIR: Circulator, EDFA: Erbium-doped fiber amplifier, LD: Laser diode, PC: Polarization controller. (b) Output power of laser diode as a function of driving current. (c) Optical spectrum of the laser diode output.

In experiments, the impact of random distributed feedback from the fiber random grating on the chaotic output of the slave LD was firstly investigated. To this end, the master LD was disconnected from injecting laser output into the slave LD. The driving current for the slave LD was set to around 6.5mA, which is slightly above the threshold value. With this driving current, the slave LD emits light with a power less than 100μW. After being amplified by the EDFA, the intra-cavity light is randomly reflected by the large numbers of scattering centers along the fiber random grating. During the experiments, the fiber random grating was put into a soundproof box to be isolated from environmental perturbations. Since the backscattering from the fiber random grating is sensitive to the SOP of the input light [31], the SOP of the intra-cavity light is adjusted by PC 1 to optimize the chaotic laser bandwidth. Fig. 2 shows the experimentally obtained evolution of the optical spectra and radio-frequency spectra of the chaotic slave LD when only the random distributed feedback from the fiber random grating is available. The delay length in the chaotic laser system is around 63.28m, which corresponds to the feedback roundtrip delay time of 316.4ns. The feedback strengths are varied by the optical attenuator from -45dBm to -30dBm. The optical spectra in Fig. 2 clearly show that the linewidth of the slave LD is broadening when the random distributed feedback strength is increasing. Meanwhile the radio-frequency spectra of the chaotic laser reveal that the bandwidth of the chaotic output is increased from 2.25GHz to 3.55GHz when the feedback strength is enhanced to -30dBm. Above this feedback strength, the bandwidth of the chaotic output remains almost unchanged, which is due to the limited gain provided by the slave laser.

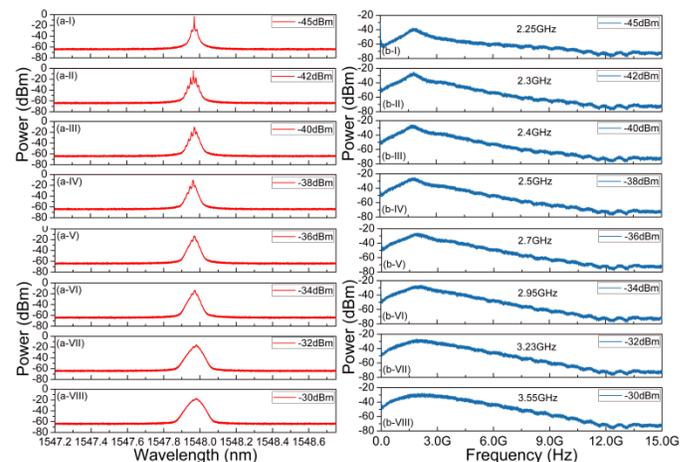

Fig. 2. Experimentally obtained evolution of the optical spectra (a) and radio-frequency spectra (b) of the slave LD only subjected to the random distributed feedback from the fiber random grating with feedback strengths ranging from -45dBm to -30dBm.

In order to further enhance the bandwidth of the chaotic output, light from the master LD is injected into the slave LD as shown in Fig. 1(a). The power of the laser injection is controlled by the 2nd optical attenuator. The SOP of the injected light is adjusted by PC 2 so as to optimize the bandwidth broadening. Fig. 3(a) shows the optical spectra of the slave LD when it is free running, only subjected to the optical feedback, and subjected to both the optical feedback and laser injection. These spectra were measured with optical feedback strength of -30dBm, injection strength of -5.5dBm and injection frequency difference of 1.45GHz between the master and slave lasers. It is observed that the original spectrum of the slave laser undergoes red-shift induced by the change of carrier density. In Fig. 3(b), the radio-frequency spectra for different conditions are illustrated, where the bandwidth enhancement of the power spectra of the chaotic signal is observed. The pink and black lines denote the power spectra of the free-running slave LD and the photo-detector (PD) noise floor, respectively. The small peak shown in the pink curve corresponds to the relaxation oscillation of the free-running LD. When the slave LD is subjected to the optical feedback from the fiber random grating and injection from the master LD, the bandwidth of the power spectra is significantly broadened due to the coherence collapse effect in the semiconductor laser. Many extended cavity modes are excited in the laser system and the slave laser evolves into chaotic oscillations with broadened frequency components as shown in the blue curve of Fig. 3(b). Meanwhile, the injected light from the master LD interacts with light from the slave LD as well as the optical feedback, making the gain-coupled laser system work at unlocked state and leading to high-frequency oscillations with broad bandwidth. We define the bandwidth of chaotic signals as the span between the DC and the frequency where 80% of the energy is contained [32]. As shown in the red curve in Fig. 3(b) where the slave LD is only subjected to optical feedback, the bandwidth of the chaotic output is measured to be 3.55GHz and the spectrum decays rapidly to the noise floor beyond 15GHz. The blue curve shows the measured power spectrum when the slave LD is subjected to both the optical feedback and laser injection. With optimization of the detuning frequency between the slave LD and the master LD, the maximum bandwidth of the chaotic signal achieved is measured to be 8.5GHz, which is more than twice that of the case only with optical feedback.

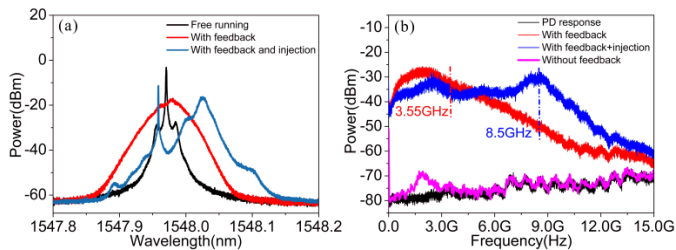

Fig. 3. Comparison of optical spectra (a) and radio-frequency spectra (b) of the slave LD when subjected to the following three conditions: i) free running; ii) feedback only; iii) feedback and laser injection.

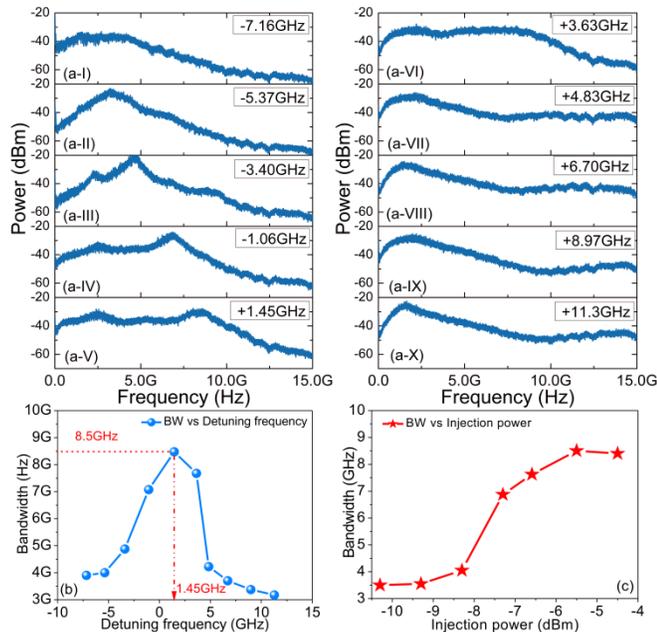

Fig. 4. (a) Experimentally obtained radio-frequency spectra of the chaotic laser output with different detuning frequencies between the slave LD and master LD. (b) Effective bandwidth of the chaotic laser as a function of detuning frequency. (c) The dependence of the effective bandwidth on laser injection power.

Fig. 4(a) shows the evolution of the radio-frequency spectra of the chaotic laser as a function of the detuning frequency. The frequency difference between the master LD and the slave LD is varied from -7.16GHz to +11.3GHz. The injection laser power from the master LD is kept around -5.5dBm and the feedback strength is kept at -30dBm. It is clearly observed that the power spectra of the chaotic signals undergo significant broadening when the detuning frequencies are small. The effective bandwidth of the chaotic laser as a function of the detuning frequency is summarized in Fig. 4(b). It is illustrated that when the detuning frequency is around +1.45GHz, the chaotic laser output exhibits the maximum bandwidth around 8.5GHz. When the frequency difference between the two LDs becomes larger, the bandwidth of the chaotic signals starts to decrease. This phenomenon further confirms that the bandwidth-enhanced chaotic laser is mainly due to the beat effect between the injected laser and the chaotic slave laser field. The chaotic bandwidth can only be maximized when the injected light is detuned to the spectral edge of the chaotic slave laser field and the beating frequency exceeds the original laser bandwidth. When the detuning frequency becomes larger, the excited high-frequency oscillations will be gradually decoupled from the original chaotic oscillations, resulting in the cease of the bandwidth broadening as the gain-coupled state is lost.

The dependence of the effective bandwidth on laser injection power is also investigated in experiments as shown in Fig. 4(c). The detuning frequency between the two LDs is set to +1.45GHz. In this condition, it is experimentally found that the bandwidth of the chaotic laser could be enhanced from 3.55GHz to 8.5GHz with the injection laser power varying from -10.3dBm to -5.5dBm. When the injection power is further increased, the broadening of the bandwidth of the chaotic output stops due to the transition of the oscillation state of the slave LD from chaotic oscillation to quasi-periodic oscillation and other possible states.

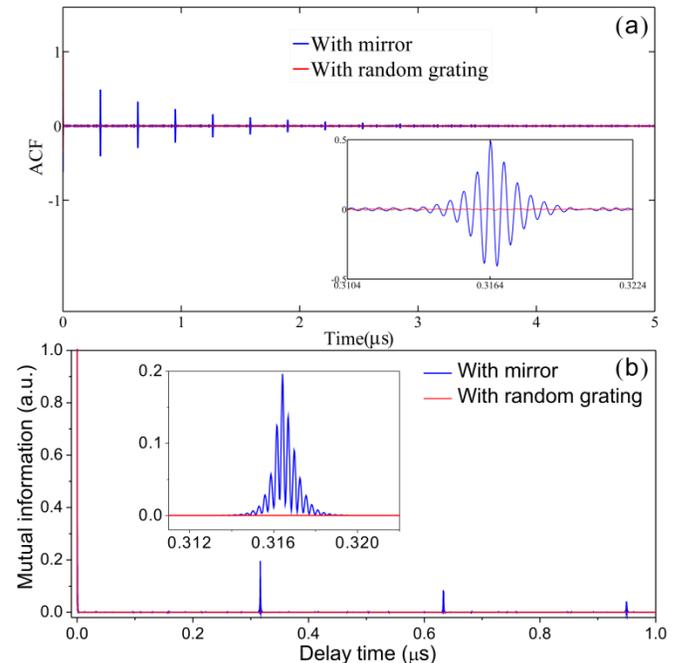

Fig. 5. ACFs (a) and mutual information (b) of the chaotic laser output with mirror feedback (blue) and random grating feedback (red). Insets: enlarged ACF and mutual information results around the delay time originated from the optical feedback.

To investigate the TDS of the chaotic output, auto-correlation function (ACF) calculations are performed on two time series signals when the slave LD is subjected to the optical feedback from the mirror and fiber random grating, respectively. The feedback strengths of the mirror and the random grating were both set at -30dBm in the experiments. The mirror feedback configuration is equivalent to the optical feedback with a fiber ring cavity [28]. The time series of the chaotic lasing output used for calculation is recorded with a time period of 25μs and the results are shown in Fig. 5(a). In the ACF result of the case with mirror feedback as shown in the blue line, TDS peaks emerge at the delay time of 316.4ns and its integer multiples. It could be calculated that the delay length in the chaotic laser system is around 63.28m. The TDS in random bit generations can be detected by an eventual eavesdropper directly using a photodiode, which makes the generated random number reproducible and thus it is harmful in security information communications. For comparison, the ACF calculation of the chaotic time series from the case with random grating feedback shows that the TDS peaks are completely concealed (TDS value: 0.005) as shown in the inset of Fig. 5(a), which indicates that the random distributed feedback from the fiber random grating could effectively eliminate the periodicity in the time series, ensuring the randomness of the random bits to be generated. Mutual information was also computed for the chaotic laser output with mirror feedback and random grating feedback to justify the TDS concealment as shown in Fig. 5(b). It is clearly illustrated that the TDS is effectively cancelled with random grating feedback. The physics behind this phenomenon is explained as follows: as the TDS in the case of a single mirror feedback originates from the beating between the longitudinal cavity modes, the large numbers of randomly spaced scattering centers with low phase correlation in the fiber random grating introduce large numbers of uncorrelated modes competing for the limited optical gain. Chaotic lasing output is then expected. As a result, a strong and stable beating between any two of these longitudinal cavity modes is prevented.

3. Discussion and conclusion

The maximum broadened bandwidth of the current chaotic laser system is mainly limited by the relatively low relaxation oscillation frequency (~2GHz) of the LDs. This restriction could be easily overcome by using LDs or DBF lasers with high relaxation oscillation frequencies. Increasing the number of the injected lasers would be another alternative to enhance the bandwidth broadening. Note that the random distributed feedback from the fiber random grating plays an important role in concealing the time-delay signature of the chaotic output. Since the local birefringence of the randomly distributed index modification spots along the grating length varies from each other, the SOP of light that is to be reflected by the grating should be carefully adjusted so that each scattered light from the scattering centers will have contributions to increase the random cavity modes and to reduce the laser coherence.

In conclusion, a TDS-free broadband chaotic laser system has been proposed and experimentally realized with a bandwidth up to 8.5GHz. The chaotic bandwidth broadening is induced by the coupling among the optical feedback from the fiber random grating, the laser injection from the master LD as well as the original light from the slave LD. The random distributed feedback from the large numbers of scattering centers along the grating length significantly increase the complexity of the external feedback cavity structures, leading the complete suppression of the TDS of the chaotic signals. This broadband chaotic laser could be potentially developed to high-speed physical random number generators, which will find important applications in practical real-time security communication systems.

**Acknowledgments:** This work was supported by the Natural Sciences and Engineering Research Council of Canada (NSERC) Discovery Grants (RGPIN-2015-06071), the NSERC Engage Grants (506792/2016) and the NSERC RTI Grants (RTI-2017-00631).

**Conflicts of Interest:** The authors declare no conflict of interest.

**References.**
1. M. Sciamanna and K. A. Shore, "Physics and applications of laser diode chaos," Nature Photon. 9, 151 (2015).
2. A. Argyris, D. Syvridis, L. Larger, V. Annovazzi-Lodi, P. Colet, I. Fischer, J. Garcia-Ojalvo, C. R. Mirasso, L. Pesquera, and K. A. Shore, "Chaos-based communications at high bit rates using commercial fibre-optic links," Nature 438, 343 (2005).
3. A. Argyris, M. Hamacher, K. Chlouverakis, A. Bogris, and D. Syvridis, "Photonic integrated decice for chaos applications in communications," Phys. Rev. Lett. 100, 194101 (2008).
4. Y. Takiguchi, K. Ohyagi, and J. Ohtsubo, "Bandwidth-enhanced chaos synchronization in strongly injection-locked semiconductor lasers with optical feedback," Opt. Lett. 28, 319 (2003).
5. N. Q. Li, H. Susanto, B. Cemlyn, I. D. Henning, and M. J. Adams, "Secure communication systems based on chaos in optically pumped spin-VCSELs," Opt. Lett. 42, 3494 (2017).
6. A. Uchida, K. Amano, M. Inoue, K. Hirano, S. Naito, H. Someya, I. Oowada, T. Kurashige, M. Shiki, S. Yoshimori, K. Yoshimura, and P. Davis, "Fast physical random bit generation with chaotic semiconductor lasers," Nature Photon. 2, 728 (2008).
7. I. Reidler, Y. Aviad, M. Rosenbluh, and I. Kanter, "Ultrahigh-speed random number generation based on a chaotic semiconductor laser," Phys. Rev. Lett. 103, 024102 (2009).
8. A. Wang, L. Wang, P. Li, and Y. Wang, "Minimal-post-processing 320-Gbps true random bit generation using physical white chaos," Opt. Express 25, 3153 (2017).
9. P. Li, Y. Sun, X. Liu, X. Yi, J. Zhang, X. Guo, Y. Guo, and Y. Wang, "Fully photonics-based physical random bit generator," Opt. Lett. 41, 3347 (2016).
10. N. Oliver, M. C. Soriano, D. W. Sukow, and I. Fischer, "Dynamics of a semiconductor laser with polarization-rotated feedback and its utilization for random bit generation," Opt. Lett. 36, 4632 (2011).
11. L. Zhang, B. Pan, G. Chen, L. Guo, D. Lu, L. Zhao, and W. Wang, "640-Gbit/s fast physical random number generation using a broadband chaotic semiconductor laser," Sci. Rep. 7, 45900 (2017).
12. N. Q. Li, B. Kim, V. N. Chizhevsky, A. Locquet, M. Bloch, D. S. Citrin, and W. Pan, "Two approaches for ultrafast random bit generation based on the chaotic dynamics of a semiconductor laser," Opt. Exp. 22, 6634 (2014).
13. F. Y. Lin and J. M. Liu, "Chaotic lidar," IEEE J. Sel. Top. Quantum Electron. 10, 991 (2004).
14. F. Y. Lin and J. M. Liu, "Chaotic radar using nonlinear laser dynamics," IEEE J. Quantum Electron. 40, 815 (2004).
15. D. Rontani, D. Choi, C-Y. Chang, A. Locquet, and D. S. Citrin, "Compressive sensing with optical chaos," Sci. Rep. 6, 35206 (2016).
16. Y. Wang, B. Wang, and A. Wang, "Chaotic correlation optical time domain reflectometer utilizing laser diode," IEEE Photon. Technol. Lett. 20, 1636 (2008).

17. A. Wang, N. Wang, Y. Yang, B. Wang, M. Zhang, and Y. Wang, "Precise fault location in WDMPON by utilizing wavelength tunable chaotic laser," J. Lightwave Technol. 30, 3420 (2012).
18. N. Q. Li, H. Susanto, B. R. Cemlyn, I. D. Henning, and M. J. Adams, "Stability and bifurcation analysis of spin-polarized vertical-cavity surface-emitting lasers," Phys. Rev. A 96, 013840 (2017).
19. T. B. Simpson, J. M. Liu, K. F. Huang, and K. Tai, "Nonlinear dynamics induced by external optical injection in semiconductor lasers," Quantum Semiclass Opt. 9, 765 (1997).
20. M. C. Chiang, H. F. Chen, and J. M. Liu, "Experimental synchronization of mutually coupled semiconductor lasers with optoelectronic feedback," IEEE J. Quantum Electron. 41, 1333 (2005).
21. J. Mork, B. Tromborg, and J. Mark, "Chaos in semiconductor lasers with optical feedback: Theory and experiment," IEEE J. Quantum Electron. 28, 93 (1992).
22. D. Rontani, A. Locquet, M. Sciamanna, and D. S. Citrin, "Loss of time-delay signature in the chaotic output of a semiconductor laser with optical feedback," Opt. Lett. 32, 2960 (2007).
23. J.-G. Wu, G.-Q. Xia, and Z.-M. Wu, "Suppression of time delay signatures of chaotic output in a semiconductor laser with double optical feedback," Opt. Express 17, 20124 (2009).
24. S. S. Li and S. C. Chan, "Chaotic time-delay signature suppression in a semiconductor laser with frequency-detuned grating feedback," IEEE J. Sel. Topics Quantum Electron., 21, 541 (2015).
25. S. S. Li, Q. Liu, and S. C. Chan, "Distributed feedbacks for time-delay signature suppression of chaos generated from a semiconductor laser," IEEE Photon. J., 4, 1930 (2012).
26. N. Q. Li, W. Pan, S. Y. Xiang, L. S. Yan, B. Luo, and X. H. Zou, "Loss of time delay signature in Broadband cascade-coupled semiconductor lasers," IEEE Photon. Technol. Lett. 24, 2187 (2012).
27. S. Sunada, T. Harayama, K. Arai, K. Yoshimura, P. Davis, K. Tsuzuki, and A. Uchida, "Chaos laser chips with delayed optical feedback using a passive ring waveguide," Opt. Express 19, 5713 (2011).
28. A. B. Wang, Y. C. Wang, and J. F. Wang, "Route to broadband chaos in a chaotic laser diode subject to optical injection," Opt. Lett. 34, 1144 (2009).
29. M. J. Zhang, T. G. Liu, P. Li, A. B. Wang, J. Z. Zhang, and Y. C. Wang, "Generation of broadband chaotic laser using dual-wavelength optically injected Fabry–Perot laser diode with optical feedback," IEEE Photon. Technol. Lett. 23, 1872 (2011).
30. Y. Xu, P. Lu, S. Gao, D. Xiang, P. Lu, S. Mihailov, and X. Bao, "Optical fiber random grating-based multiparameter sensor," Opt. Lett. 40, 5514 (2015).
31. Y. Xu, M. Zhang, L. Zhang, P. Lu, S. Mihailov, and X. Bao, "Time-delay signature suppression in a chaotic semiconductor laser by fiber random grating induced random distributed feedback," Opt. Lett. 42, 4107 (2017).
32. F. Y. Lin and J. M. Liu, "Nonlinear dynamical characteristics of an optically injected semiconductor laser subject to optoelectronic feedback," Opt. Commun. 221, 173 (2003).